\documentclass[aps,prb,twocolumn,superscriptaddress,showpacs,showkeys,intlimits,amsmath,amssymb,floats,floatfix]{revtex4}

\usepackage{bm}
\usepackage[pdftex]{graphicx}
\usepackage{epsfig}
\usepackage[thinspace]{SIunits}
\usepackage{color}
\usepackage{amsmath}

\newcommand{\vk}{\textbf{k}}
\newcommand{\vq}{\textbf{q}}
\newcommand{\vQ}{\textbf{Q}}

\newcommand{\wn}{\mbox{$\mathrm{\,cm^{-1}{}}$}}            
\newcommand{\ErTe}{\mbox{$\mathrm{ErTe_3}{}$}}             

\newcommand{\Tone}{\mbox{$T_\mathrm{CDW1}$}}               
\newcommand{\Ttwo}{\mbox{$T_\mathrm{CDW2}$}}               

\begin{document}

\title{Alternative route to charge density wave formation in multiband systems}

\author{H.-M. Eiter}
\affiliation{Walther Meissner Institut, Bayerische Akademie der Wissenschaften, 85748 Garching, Germany}

\author{M. Lavagnini}
\affiliation{Walther Meissner Institut, Bayerische Akademie der Wissenschaften, 85748 Garching, Germany}

\author{R. Hackl$^\ast$}
\affiliation{Walther Meissner Institut, Bayerische Akademie der Wissenschaften, 85748 Garching, Germany}

\author{E.A. Nowadnick}
\affiliation{Stanford Institute for Materials and Energy Sciences, SLAC National Accelerator Laboratory, 2575 Sand Hill Road, Menlo Park CA 94025, USA}
\affiliation{Geballe Laboratory for Advanced Materials, Stanford University, Stanford CA 94305, USA}
\affiliation{Department of Physics, Stanford University, Stanford CA 94305, USA}

\author{A.F. Kemper}
\affiliation{Stanford Institute for Materials and Energy Sciences, SLAC National Accelerator Laboratory, 2575 Sand Hill Road, Menlo Park CA 94025, USA}
\affiliation{Geballe Laboratory for Advanced Materials, Stanford University, Stanford CA 94305, USA}

\author{T.P.~Devereaux}
\affiliation{Stanford Institute for Materials and Energy Sciences, SLAC National Accelerator Laboratory, 2575 Sand Hill Road, Menlo Park CA 94025, USA}
\affiliation{Geballe Laboratory for Advanced Materials, Stanford University, Stanford CA 94305, USA}

\author{J.-H. Chu}
\affiliation{Stanford Institute for Materials and Energy Sciences, SLAC National Accelerator Laboratory, 2575 Sand Hill Road, Menlo Park CA 94025, USA}
\affiliation{Geballe Laboratory for Advanced Materials, Stanford University, Stanford CA 94305, USA}
\affiliation{Department of Applied Physics, Stanford University, Stanford CA 94305, USA}

\author{J.G. Analytis}
\affiliation{Stanford Institute for Materials and Energy Sciences, SLAC National Accelerator Laboratory, 2575 Sand Hill Road, Menlo Park CA 94025, USA}
\affiliation{Geballe Laboratory for Advanced Materials, Stanford University, Stanford CA 94305, USA}
\affiliation{Department of Applied Physics, Stanford University, Stanford CA 94305, USA}

\author{I.R. Fisher}
\affiliation{Stanford Institute for Materials and Energy Sciences, SLAC National Accelerator Laboratory, 2575 Sand Hill Road, Menlo Park CA 94025, USA}
\affiliation{Geballe Laboratory for Advanced Materials, Stanford University, Stanford CA 94305, USA}
\affiliation{Department of Applied Physics, Stanford University, Stanford CA 94305, USA}

\author{L. Degiorgi}
\affiliation{Laboratorium f\"ur Festk\"orperphysik, ETH - Z\"urich, 8093 Z\"urich, Switzerland}

\date{\today}


\begin{abstract}
Charge and spin density waves, periodic modulations of the electron and magnetization densities, respectively, are among the most abundant and non-trivial low-temperature ordered phases in condensed matter. The ordering direction is widely believed to result from the Fermi surface topology. However, several recent studies indicate that this common view needs to be supplemented. Here, we show how an enhanced electron-lattice interaction can contribute to or even determine the selection of the ordering vector in the model charge density wave system ErTe$_3$. Our joint experimental and theoretical study allows us to establish a relation between the selection rules of the electronic light scattering spectra and the enhanced electron-phonon coupling in the vicinity of band degeneracy points. This alternative proposal for charge density wave formation may be of general relevance for driving phase transitions into other broken-symmetry ground states, particularly in multiband systems such as the iron based superconductors.
\end{abstract}



\maketitle

\section{Introduction}

The common view of charge density wave {(CDW)} formation was originally posed in the work by Kohn \cite{Kohn:1959}. Using Kohn's reasoning \cite{Kohn:1959}, the tendency towards ordering is particularly strong in low dimensions, because the Fermi surface has parallel parts, referred to as nesting. This nesting leads to a divergence in the Lindhard susceptibility, determining the magnitude and direction of the ordering vector \vQ \cite{Gruner:1994}. This divergence in the electronic susceptibility is conveyed to the lattice via the electron-phonon coupling, so that a phonon softens to zero frequency at \vQ\, and a static lattice distortion develops when the system enters the CDW state; a behavior known as the Kohn anomaly.

However, several  publications raise the question as to whether nesting alone is sufficient to explain the observed ordering direction \vQ~\cite{Mcmillan:1977,Varma:1983,Yao:2006,Kiss:2007,Johannes:2008}, particularly in dimensions higher than one. 
A central question is whether the selection of the CDW ordering vector is always driven by an electronic instability, or if the ordering vector could instead be determined by a lattice distortion driven by some other mechanism exploiting the role of the electron-phonon coupling. In the latter case, the selected ordering vector would not necessarily nest the Fermi surface. The importance of strongly momentum dependent electron-phonon coupling on CDW formation was pointed out in Refs. \onlinecite{Mcmillan:1977} and \onlinecite{Varma:1983}, where the relevance of the Fermi surface for determining the ordering vector was indeed found to decrease as the coupling strength increases.  In a recent paper on inelastic x-ray scattering measurements on 2$H$-NbSe$_2$, acoustic phonons were observed to soften to zero frequency over an extended region around the CDW ordering vector \cite{Weber:2011}. The authors argue that this behavior is not consistent with a Kohn anomaly picture, where sharp dips are expected. Therefore, the phonon softening must be driven by another mechanism, which they identify as a wavevector-dependent electron-phonon coupling. In addition, previous studies on chromium~\cite{Lamago:2010} and ruthenium~\cite{Heid:2000} have also shown dips in phonon dispersions arising from such anisotropic electron-phonon matrix elements.

%
\begin{figure}[h]
  \centering
  \includegraphics[width=0.45\textwidth]{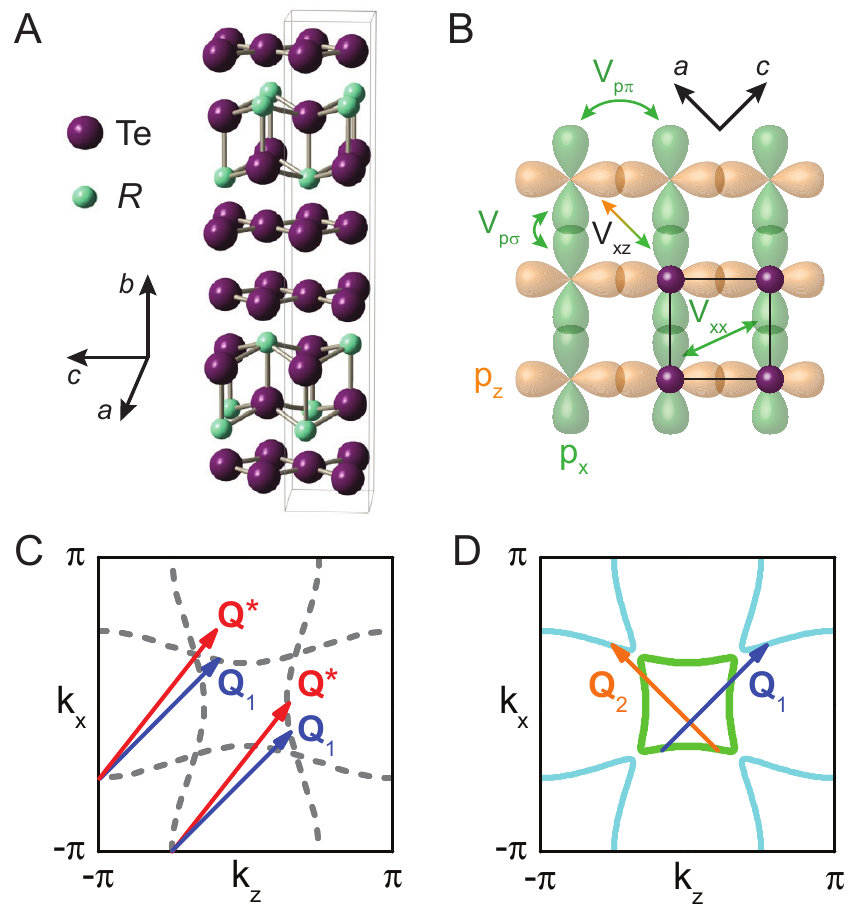}
  \caption{Real and reciprocal space structure of rare earth tri-tellurides. (\emph{A}) Crystal structure with violet and green spheres representing tellurium and rare earth atoms, respectively. The 3D crystallographic cell is indicated in grey. Note that the Te planes are perpendicular to the $b$ axis \cite{Ru:2006}. (\emph{B}) Orbital character of the Te\,$5p$ orbitals in the Te plane. The unit cell is indicated by a full line. The band structure near the Fermi energy $E_{\rm F}$ is derived from the Te\,$5p$ orbitals. The various hopping matrix elements are indicated. (\emph{C}) Fermi surface as derived from the $p_x$ and $p_z$ orbitals alone, ($V_{p\sigma}=2.99$, $V_{p\pi}=-1$, $V_{xx}=0.09$\,eV, and $V_{xz}=0$, for details see SI). Note that only $V_{xz}$ contributes to the hybridization. There are two energetically possible orientations for CDW ordering, $\vQ_1$ and $\vQ^{\ast}$, where $\vQ^{\ast}$ is the wavevector predicted by nesting. (\emph{D}) Theoretical Fermi surface for $V_{xz}=0.12$\,eV best reproducing the experimental findings \cite{Brouet:2008}. Also shown are the two experimentally observed orthogonal ordering vectors \vQ$_1$ and \vQ$_2$ parallel to the crystallographic $c$ and $a$ axes, respectively, corresponding to the CDW transitions at $T_{\rm CDW1}$ and $T_{\rm CDW2}$.}
  \label{fig:structure}
\end{figure}
%

For exploring a possible relation between anisotropic electron-phonon coupling and CDW ordering selection, it is desirable to map out the coupling strength in momentum space. For certain types of phonons, the electron-phonon matrix element is proportional to the electron-photon matrix element from Raman spectroscopy. As a result, Raman spectroscopy, which efficiently projects out different regions of the Brillouin zone with different photon polarizations, can provide an indirect method for investigating the momentum-dependence of the electron-phonon coupling in a system.

To set the stage for our discussion, we introduce the two-dimensional rare-earth tri-tellurides ($R$Te$_3$), as shown in Fig.\;\ref{fig:structure}\emph{A}. Among them the prototypical ErTe$_3$ (see Appendix~\ref{appendixA}) undergoes a first CDW transition at \Tone\ $=265$\,K, followed by a second one at \Ttwo\ $=155$\,K and allows a robust access to its intrinsic CDW properties. The ordering vectors $\vQ_{1}$ and $\vQ_{2}$ are parallel to but incommensurate with the reciprocal lattice vectors $c^{\ast} \parallel c$ and $a^{\ast} \parallel a$, respectively~\cite{Ru:2006,Brouet:2008,Ru:2008a,Yusupov:2008,Schmitt:2008,Lavagnini:2008,Lavagnini:2010,Pfuner:2010,Lazarevic:2011,Hu:2011}. The electronic properties of these layered CDW compounds can be modeled by considering a single Te plane [Fig.\;\ref{fig:structure}\emph{B} and Appendix~\ref{appendixC}:\,Supporting Information~(SI)]. The two dominant hopping terms are $V_{p\sigma}$ and $V_{p\pi}$ along and  perpendicular to the overlapping $p_x$ or $p_z$ orbitals, respectively, leading to slightly warped Fermi surface planes (Fig.\;\ref{fig:structure}\emph{C}). For this band structure, Yao and coworkers \cite{Yao:2006} studied the influence of band-filling and electron-phonon coupling strength on the charge ordering and established a strong coupling limit for the experimentally observed stripe-like CDW state. Additionally, Johannes and Mazin \cite{Johannes:2008} found that the Lindhard susceptibility has  peaks of comparable size at the nesting vector $\vQ^{\ast}$ predicted from the band structure, and at the CDW ordering vector $\vQ_1$, determined experimentally in $R$Te$_3$ \cite{Ru:2008a} (Fig.\;\ref{fig:structure}\emph{C}). This is quite similar to a recent observation in NbSe$_2$ \cite{Kiss:2007}. The inclusion of a small but non-zero hopping term between $p_x$ and $p_z$, $V_{xz} \ne 0$, lifts the degeneracy at the intersection points of the two quasi 1D Fermi surfaces, as emphasized in Fig.\;\ref{fig:structure}\emph{D}, and improves the agreement with the experimental Fermi surface, even though $\vQ^\ast$ remains the best nesting vector \cite{Brouet:2008}. Therefore, mechanisms beyond purely electronic ones have been conjectured to play an important role in selecting the ordering vector and in density wave formation. These include orthorhombicity, the tendency towards phase separation and nematicity via the Coulomb interaction, strongly momentum dependent electron-phonon interaction due to peculiarities of the band structure, breakdown of the Coulomb screening and other competing instabilities, such as magnetism \cite{Yao:2006,Johannes:2008,Emery:1990,Marder:1990,Grilli:1991,Kivelson:2003}.

In this Research Report, we analyze data from Raman experiments and the related selection rules for ErTe$_{3}$ and demonstrate that the lifting of band degeneracies enhances the light-scattering sensitivity and, concomitantly, the electron-phonon coupling at ordering vectors that do not coincide with those vectors predicted by nesting alone.

\section{Results}

\subsection{Fluctuation regime above the CDW transition temperature} %

\begin{figure}
  \centering
  \includegraphics[width=0.45\textwidth]{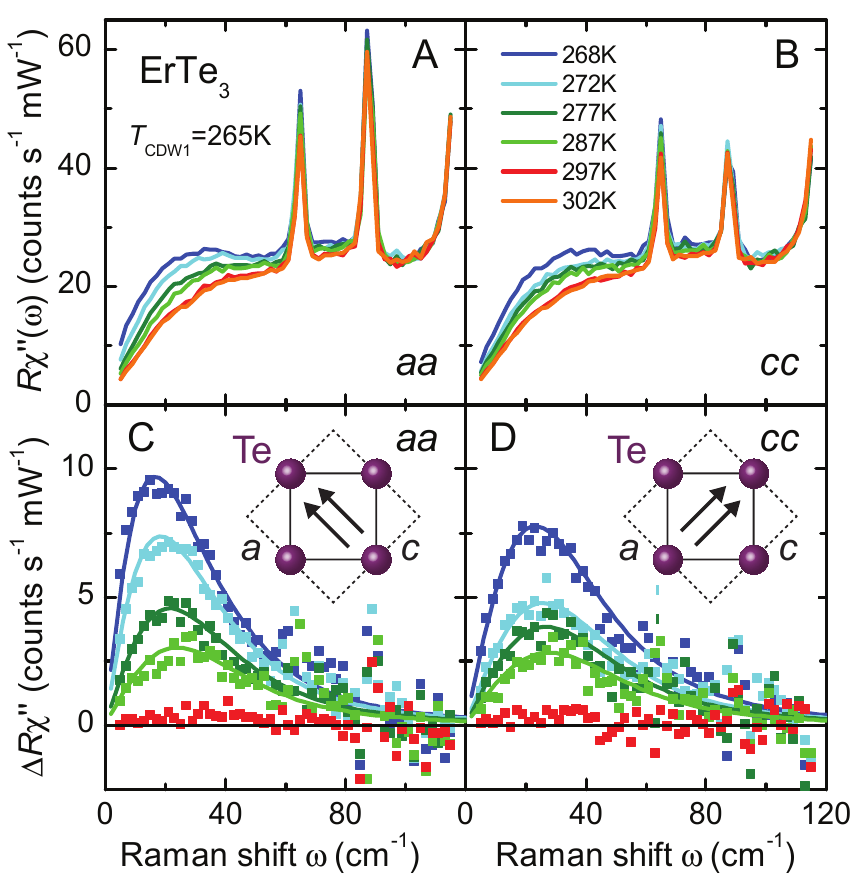}
  \caption{Normal state Raman scattering response of ErTe$_3$ at low energies. The imaginary part $R\chi^{\prime\prime}_{i,s}(\omega,T)$ of the Raman response is shown in panel (\emph{A}) for \textit{aa}- and in panel (\emph{B}) for \textit{cc}-configuration with the first and the second label representing the polarizations {\bf e}$_i$ and {\bf e}$_s$ of the incoming and, respectively, scattered photons as indicated in the insets of panels (\emph{C}) and (\emph{D}). The spectra display the presence of a fluctuation-induced response for a temperature range of about 30\,K above \Tone. In \textbf{c} and \textbf{d} the fits after Ref.~
  \cite{Caprara:2005} to the fluctuation contribution $\Delta R \chi^{\prime\prime}=R \chi^{\prime\prime}(\omega,T)-R \chi^{\prime\prime}(\omega,302\,{\rm K})$ are shown.}
  \label{fig:fluctuations}
\end{figure}

We first display the low frequency Raman spectra above \Tone~ in Fig.\;\ref{fig:fluctuations}. The narrow lines superposed on the continuum are the Raman-active phonons of the high-temperature phase \cite{Lavagnini:2008}. Instead of the expected flat continuum \cite{Kostur:1992}, strongly temperature dependent shoulders, emerging from the normal metallic response, are observed for $\Tone < T < 300$\,K in the low-energy part of the spectra (Figs.~\ref{fig:fluctuations}\emph{A} and \emph{B}). These excitations have similar intensity in $aa$ and $cc$ polarization configurations (defined in the Appendix~\ref{appendixA}), soften and get stronger upon approaching \Tone\, from higher temperatures. Above 300\,K the spectra are essentially temperature independent as expected for a metal with an almost constant resistivity \cite{Ru:2008a}.

%
%

In Fig.\;\ref{fig:fluctuations}\emph{C} and \emph{D} we show the strongly temperature dependent parts of the spectra alone which closely follow the prediction of Caprara and coworkers for the exchange of fluctuations \cite{Caprara:2005}. A fluctuation regime, suppressing $T_{\rm{CDW1}}$ below the mean field transition temperature and out of which the CDW state emerges, is expected to exist at temperatures well above $T_{\rm{CDW1}}$, particularly in our case with a large ratio $2\Delta_{\rm{1}}/k_{\rm{B}} T_{\rm{CDW1}}\sim 15$ being approximately four times the canonical mean field value of 3.53 \cite{Gruner:1994}. Signatures of such CDW precursor effects were indeed observed by x-ray diffraction \cite{Ru:2008a} and by optical (IR) spectroscopy \cite{Pfuner:2010,Hu:2011}. Here, we further reveal the nature and the symmetry properties of the fluctuations; our observations are compatible with the $A_{1g}$ channel, which indicates the survival of the C$_4$ rotational symmetry of the pseudo-tetragonal phase. [The in-plane lattice parameters are almost identical; therefore, the \rm{Te}	planes are essentially square (C$_4$-symmetric), although the structure is fundamentally orthorhombic because of the glide plane between adjacent \rm{Te} layers.] The survival of the C$_4$ rotational symmetry excludes the presence of precursor effects due to nematic order, as, e.g., intensively debated in the cuprates \cite{Kivelson:2003,Benfatto:2000}, which would appear in $B_{1g}$ symmetry.

\subsection{CDW amplitude mode excitations and electron-phonon coupling strength} %

\begin{figure}[t]
  \centering
  \includegraphics[width=0.45\textwidth]{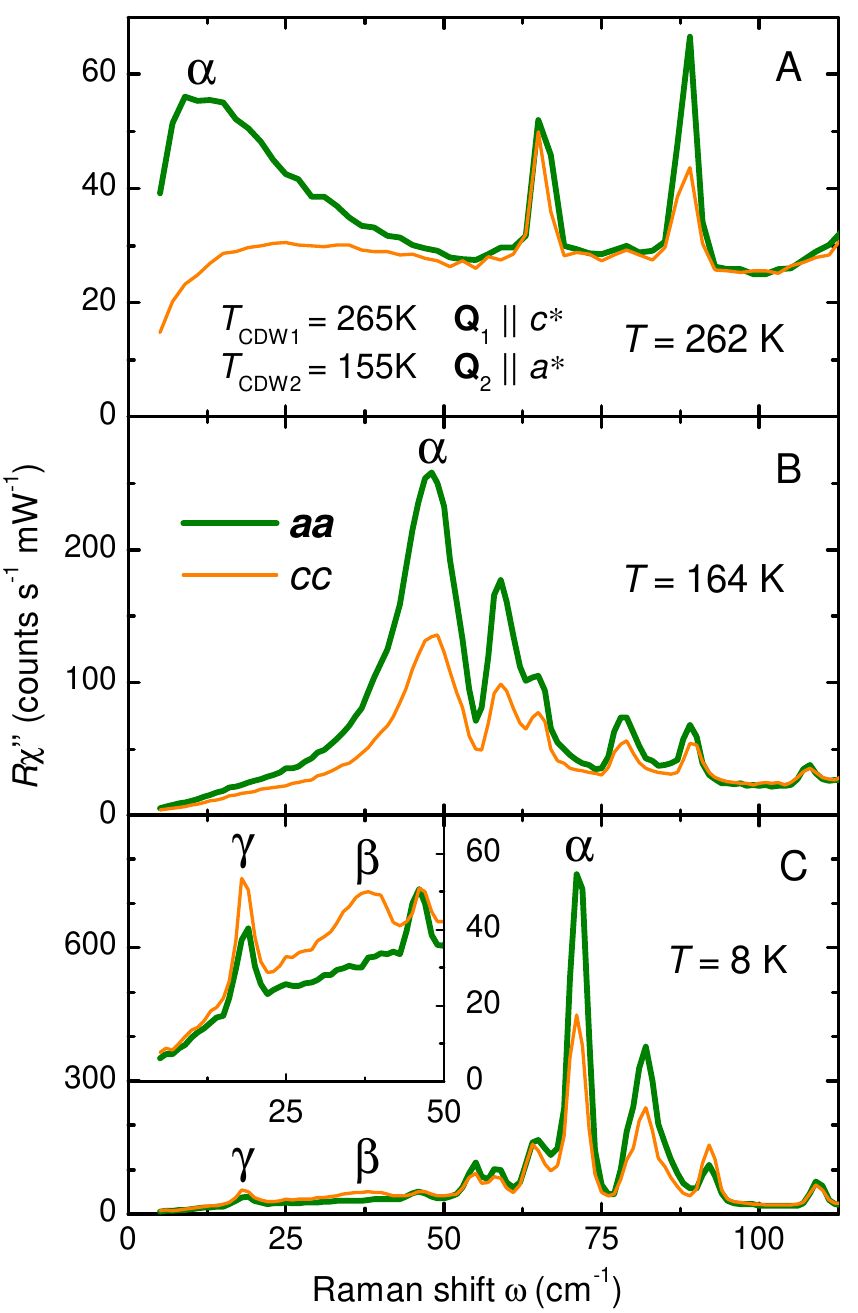}
  \caption{Amplitude modes of ErTe$_3$ for the two CDW transitions. The three panels show comparisons of the $aa$ and $cc$ polarized spectra at selected temperatures. Note the different intensity scales. $c^\ast$ and $a^\ast$ are reciprocal lattice vectors parallel to the crystallographic lattice vectors $c$ and $a$, respectively. The amplitude mode (peak\;$\alpha$) of the first transition displays a $\sim$\,2:1 intensity-ratio between the $aa$ and $cc$ polarizations, while the second AM (peak\;$\beta$) is fully $cc$ polarized. The inset in (\emph{C}) shows the second amplitude mode (peak\;$\beta$) and beat mode (peak\;$\gamma$) at 18 \wn~on an expanded intensity scale.
  }
  \label{fig:AM}
\end{figure}

Immediately below \Tone~the amplitude mode (AM) of the CDW pops up and gains intensity with decreasing temperature, shown as peak\;$\alpha$ in Fig.\;\ref{fig:AM}. The AM appears in both polarizations, $aa$ and $cc$, with an intensity ratio of $\sim$\,2:1. Upon further cooling the AM moves to higher energies, couples to phonons \cite{Lazarevic:2011}, and gains more than one order of magnitude in intensity \cite{Lavagnini:2010}. At the lowest temperature it saturates at $\omega_{\rm AM}=71\wn$ (Fig.\;\ref{fig:AM}\emph{C}, peak\;$\alpha$). The relation $\omega_{\rm AM} = \sqrt{\lambda} \omega_{2\vk_F}$ \cite{Gruner:1994} at $T=0$ between the energy of the AM  and the un-renormalized CDW phonon energy ($\omega_{2\vk_F}=110\pm20$\wn, Ref.~\onlinecite{Lavagnini:2008}) leads to $\lambda = 0.4\pm 0.1$. In contrast to superconductors $\lambda = 0.4$ is already in the strong coupling regime, since it is well beyond the threshold of 0.103, which separates nematic from stripe order \cite{Yao:2006}. As shown in the inset of Fig.\;\ref{fig:AM}\emph{C} the second AM (peak\;$\beta$) is fully $cc$ polarized and saturates at 38\wn. Another mode (peak\;$\gamma$ in Fig.\;\ref{fig:AM}\emph{C}) with resolution-limited width appears at $\Omega_{\delta} \sim 18$\wn. The full analysis yields tetragonal $B_{1g}$ symmetry, and the energy corresponds to the beat frequency of the two amplitude modes $\Omega_{\rm beat} = 1/2|\Omega_{\rm AM1}-\Omega_{\rm AM2}|$ to within the experimental error. Microscopically, the coupling between collective states may result from eigenvector mixing or because the two condensates share common parts of the Fermi surface \cite{Littlewood:1981,Browne:1983,Tutto:1992}.

%
%

\subsection{Temperature dependence and anisotropy of the CDW gap}

\begin{figure}[t]
  \centering
  \includegraphics[width=0.45\textwidth]{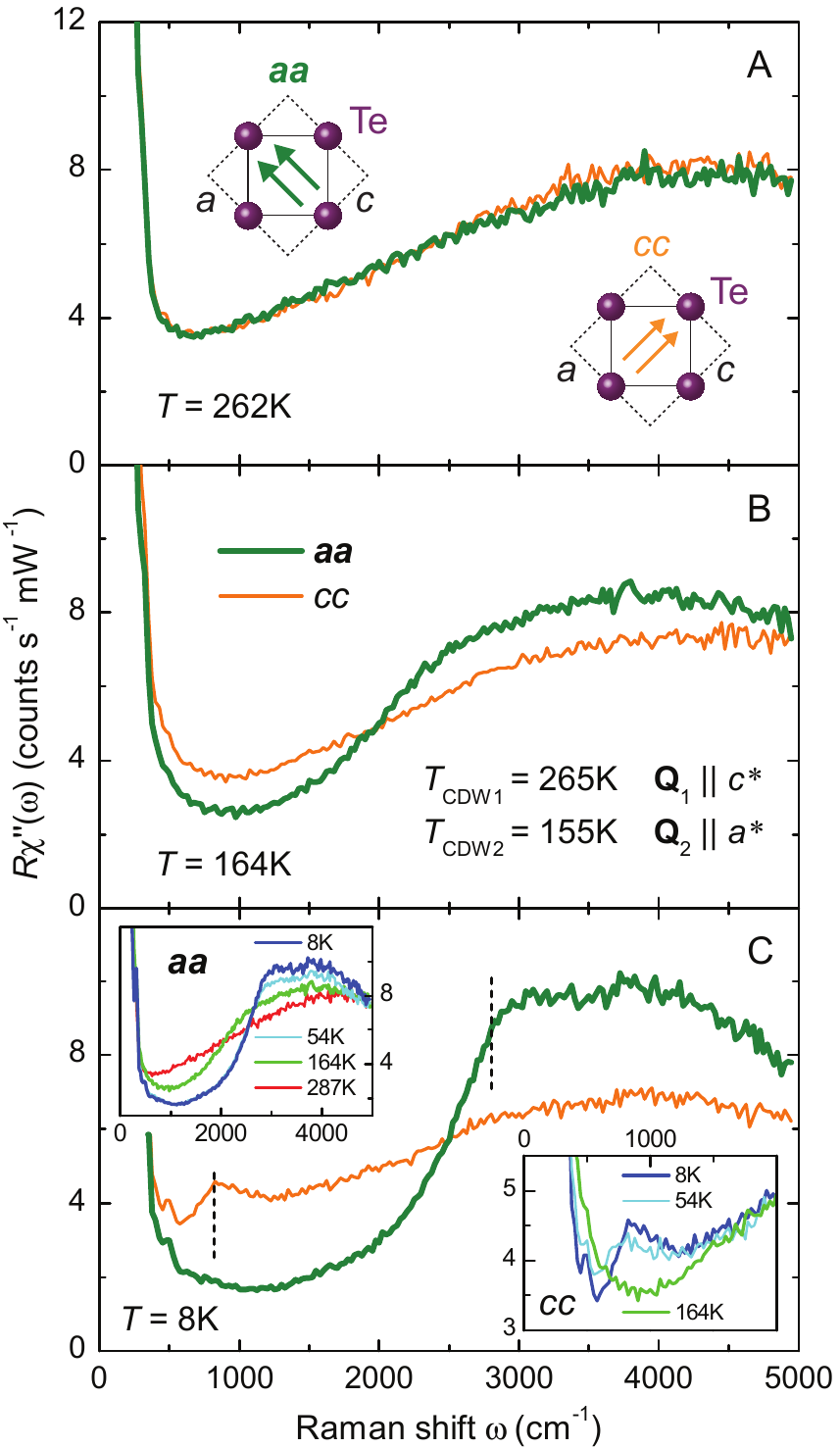}
  \caption{Temperature dependence of the high energy Raman spectra in ErTe$_3$. (\emph{A}) At 262\,K there is no difference in the electronic excitations at high energy for both polarizations. The insets sketch the incoming and scattered photon polarizations (defined in the Methods section). (\emph{B}) At 164\,K the anisotropy between the two polarizations is already well resolved. (\emph{C}) In the limit $T \to 0$ the electronic gaps with edges at 800 and 2800 \wn (dashed lines) for both CDWs are fully developed. The upper left and lower right insets show the temperature dependences of both CDW gaps, using the same colour-code. Note that the signatures of the CDW transition at  \Tone~ with $\vQ_1 \parallel c^{\ast}$ are observed with incoming and scattered light polarizations parallel to $a$. The opposite is true for the transition at \Ttwo~ as shown in detail in Fig.\;\ref{fig:theory}.
  }
  \label{fig:gap}
\end{figure}

Fig.~\ref{fig:gap} depicts the electronic Raman response of ErTe$_3$ at various temperatures below \Tone. The spectra at 262\,K (Fig.\;\ref{fig:gap}\emph{A}) are isotropic, rise almost linearly between 800 and 3500\,\wn~and finally become flat. Upon lowering the temperature there is a transfer of spectral weight in  the $aa$ spectrum from low to high energies (Fig.\;\ref{fig:gap}\emph{B} and \emph{C}). At 8\,K (Fig.\;\ref{fig:gap}\emph{C}), there is a {relatively} weak new structure in the $cc$ spectrum in the range 500-1200 \wn. The insets in Fig.\;\ref{fig:gap}\emph{C} highlight the temperature dependences of the $aa$ and $cc$ polarized spectra right above and below \Tone~ and \Ttwo, respectively. In either case, spectral weight is progressively suppressed below the gap edges and piles up above. As opposed to the AMs (Fig.\;\ref{fig:AM}), there is a full anisotropy without any leakage between the two orthogonal $aa$ and $cc$ directions, indicating that the crystal is single domain in the probed spot. Single domain areas were already observed in an earlier angle-resolved photoemission spectroscopy (ARPES) experiment on ErTe$_3$ \cite{Moore:2010}. We identify the edges with twice the maximum gap energies of the first and the second CDW, $2\Delta_1=2800$\wn\, and $2\Delta_2=800$\wn, respectively, in agreement with ARPES findings \cite{Moore:2010}.

%

\subsection{Raman selection rules and anisotropic electron-lattice coupling}

We now elaborate on the remarkable selection rules described in the preceding paragraph (Fig.\;\ref{fig:gap}) and relate them to hybridization effects of the band structure. In addition and more importantly, we demonstrate that the selection rules and the strong anisotropy of the electron-phonon coupling, which influences the CDW ordering, are intimately connected and just two sides of the same coin.

\begin{figure}
  \centering
  \includegraphics[width=0.45\textwidth]{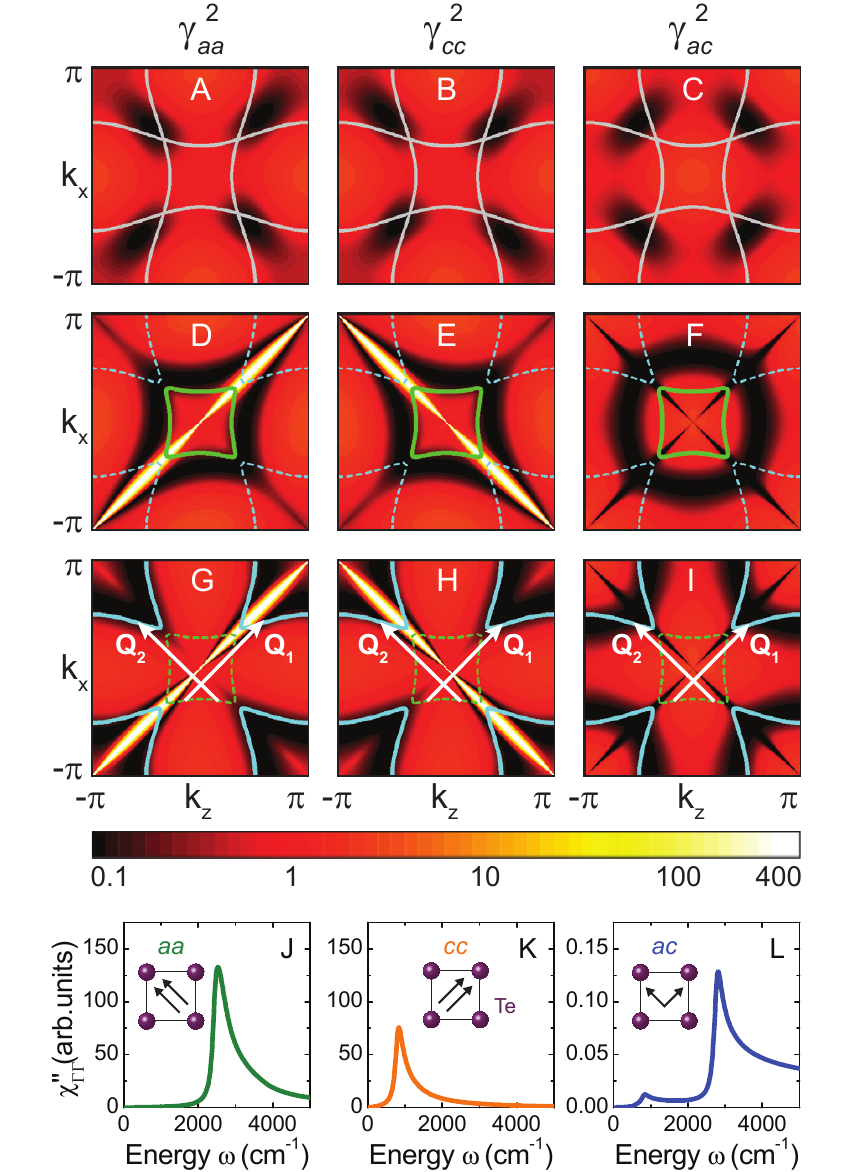}
  \caption{Theoretical prediction for the Raman vertices and spectra of ErTe$_3$. (\emph{A}-\emph{I}) All vertices are derived from the band structure according to Eq.\,(\ref{eq:gamma}) in SI. (\emph{A}-\emph{C}) The first row shows the vertices without 2D coupling ($V_{xz}=0$) and the corresponding Fermi surfaces. (\emph{D}-\emph{F}) Raman vertices for the bands corresponding to the central part of the Fermi surface (green) and (\emph{G}-\emph{I}) to the outer part (light blue). The focussing effect due to the lifted degeneracy enhances the vertices by more than two orders of magnitude as indicated by the colour code. The ordering vectors are displayed in the third row. (\emph{J}-\emph{L}) All spectra are calculated at $T<\Ttwo$ and include both CDWs (Eq.\,(\ref{eq:response}) in SI). (\emph{J}, \emph{K}) For parallel polarizations one observes only the CDW with ordering vector perpendicular to the light polarizations. The response of the respective orthogonal CDW is too weak to be visible. (\emph{L}) For $ac$ polarization both gaps can be resolved but the overall intensity is more than three orders of magnitude lower than that in the two other configurations. This is well below the detection limit, and, in fact, no signatures of the gaps can be observed experimentally in $ac$ polarization (see Fig.\;\ref{fig:nogap} in SI).
  }
  \label{fig:theory}
\end{figure}

The electronic Raman response including the selection rules can be derived directly from the band structure and the momentum dependences of the CDW gaps using the formalism of V\'anylos and Virosztek \cite{Vanyolos:2005}. The intensity of the light scattering for different polarization combinations (Raman vertices) are mainly determined by the curvature of the electronic dispersion, as described in detail in Ref.~\onlinecite{Devereaux:2007}. If we neglect hybridization ($V_{xz}=0$, see Fig.\;\ref{fig:structure}\emph{B}), the Raman vertices are almost featureless with little highlights in particular regions of the Fermi surface, as illustrated in Fig.\;\ref{fig:theory}\emph{A}-\emph{C} for the $aa$, $cc$ and $ac$ polarizations. Upon including hybridization, the band degeneracy is lifted (Fig.\;\ref{fig:structure}\emph{D}), and the two bands exhibit strong curvatures (Fig.\;\ref{fig:structure}\emph{D} and Fig.\;\ref{fig:band} in SI). As a result, the vertices become highly focused along the diagonals of the Brillouin zone for parallel polarizations, as shown in Fig.\;\ref{fig:theory}\emph{D}, \emph{E} and \emph{G}, \emph{H} for both bands, because of nearly singular band curvature \cite{Mazin:2010a}. This focusing enhances the light scattering precisely at the Fermi surface points connected by the CDW ordering wavevectors.

The focusing effect on the electronic spectra can be demonstrated directly via the weak-coupling Raman response $\chi^{\prime\prime}_{\Gamma,\Gamma}$ [see SI, Eq.\;(\ref{eq:response})]. We assume that at $T<T_{\rm CDW2}$ the two perpendicular CDWs with ordering vectors ${\bf Q}_1$ and ${\bf Q}_2$ have fully developed gaps $\Delta_1$ and $\Delta_2$, respectively. The ${\bf Q}_1$ and ${\bf Q}_2$ vectors connect the corners of the electron pocket around the $\Gamma$ point with the corners of the hole pockets (Fig.\;\ref{fig:structure} and Fig.\;\ref{fig:theory}). The response for the $aa$, $cc$, and $ac$ polarization orientations is shown in Fig.\;\ref{fig:theory}\emph{J}-\emph{L}. No mixing can be observed in the spectra with parallel polarizations (panels \emph{J} and \emph{K}). In the $ac$ configuration both gaps are in principle visible (panel \emph{L}), but the expected intensity is three orders of magnitude smaller than in $aa$ and $cc$ and cannot be observed in the experiment (see Fig.\;\ref{fig:nogap} in SI).

As we do not include any other scattering mechanisms, phase space restricts the non-resonant creation of electron-hole excitations to points where the CDW mixes particles with wavevectors $\vk$ and $\vk+\vQ$ across the Fermi surface. Consequently, light scattering is enhanced where energy is gained due to the CDW gap opening at the Fermi surface (Fig.\;\ref{fig:theory}\emph{J}-\emph{L}). Raman scattering efficiently projects out the relevant parts of the Brillouin zone in such a multiband system, so that the signal is small at low energies below the gap edge but is significantly enhanced at twice the CDW gap (Fig.\;\ref{fig:theory}\emph{J}-\emph{L}), in agreement with our experiment (Fig.\;\ref{fig:gap}\emph{C}).

%
%

\begin{figure}
  \centering
    \includegraphics[width=0.45\textwidth]{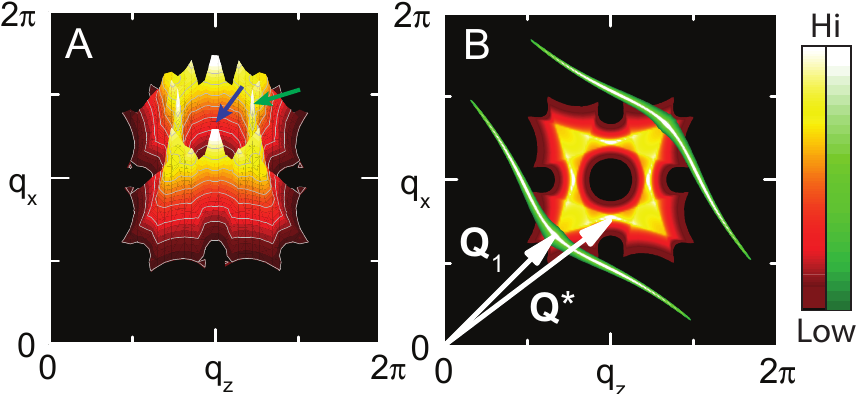}
  \caption{Comparison of the susceptibilities. (\emph{A}) 3D plot of the real part $\chi_L'$ of the Lindhard susceptibility. There is little structure around the rim. There are two orthogonal but equivalent ordering directions $\vQ^{\ast}$ as indicated by a blue and a green arrow. (\emph{B}) 2D superposition of the real parts of the Lindhard susceptibility ($\chi_L'$, yellow-red) and the projected susceptibility for interband scattering transitions ($\chi_P'$, green). For clarity, only one ordering direction is shown. It is the focusing effect of the stress tensor which selects the experimentally observed ordering wave vector $\vQ_1$.
  }
  \label{fig:phonon_and_lindhard}
\end{figure}

Obviously, the lifting of band degeneracies dramatically affects the Raman selection rules by locally enhancing the Raman vertex (Fig.\;\ref{fig:theory}) which is proportional to the inverse effective mass tensor \cite{Devereaux:2007}. Now we make use of an analogy between electron-phonon and electron-photon scattering, where the electrons scatter from a phonon rather than a photon and the electron-phonon coupling vertex replaces the Raman vertex \cite{Devereaux:1992}. In particular, for the case of stress phonons, the electron-phonon coupling vertex is given by the electronic stress tensor, which is proportional to the inverse effective mass tensor \cite{Shastry:1990,Einzel:2008,Miller:1967,Keck:1976}. As a result, for the system studied here,  the electron-phonon coupling vertex, like the Raman vertex, is highly anisotropic in momentum space. While the Lindhard susceptibility $\chi_L$ is relevant for a momentum-independent electron-phonon coupling, in this case we must include the momentum dependent electron-phonon coupling vertex into the electronic susceptibility; we call this susceptibility $\chi_P$ (see Appendix~\ref{appendixB} and Appendix~\ref{appendixC}:\,SI) \cite{Varma:1977}. Fig.\;\ref{fig:phonon_and_lindhard} illustrates the importance of including the electron-phonon coupling vertex, where we compare the real parts $\chi_L^{\prime}$ and $\chi_P^{\prime}$ of both susceptibilities. Whereas the Lindhard susceptibility $\chi_L^{\prime}$ has maxima of comparable height for several different ordering vectors (Fig.\;\ref{fig:phonon_and_lindhard}\emph{A}) and therefore does not lead to an unambiguous selection of one of them, $\chi_P^{\prime}$ contributes to the instability at the proper location in \vq~ space and finally selects the experimentally observed ordering vector \vQ$_1$ (Fig.\;\ref{fig:phonon_and_lindhard}\emph{B}). Here, $\vq$ is the difference of the momenta $\vk$ and $\vk^{\prime}$ of a scattered electron. Furthermore, as noted by Yao \textit{et al.} \cite{Yao:2006}, any enhancement of the averaged electron-phonon coupling strength $\lambda$ (definition in SI) will drive the system further towards the observed order. These two effects conspire to minimize the dependence on model details.

\section{Discussion}

%
%

The huge modulation of the Raman vertex indicates strong fluctuations of the charge density in the vicinity of the degeneracy points. These fluctuations manifest themselves in the Raman response above $\Tone$ (Fig.\;\ref{fig:fluctuations}) where the lattice still has the full C$_4$ symmetry. Yao and coworkers showed in the framework of a Landau expansion of the free energy \cite{Yao:2006} that both charge fluctuations and electron-phonon coupling cooperate to drive the system towards the experimental ordering vector. The substantial charge fluctuations couple to and soften the phonon relevant for the CDW ordering.  Taking into account the large electron-phonon coupling near the band degeneracy points (Fig.\;\ref{fig:theory}), the charge fluctuations actually renormalize the phonon frequency at the momenta $\bf{q}=\bf Q_{\rm{1}} \,(\bf Q_{\rm{2}})$ rather than \vQ$^{\ast}$ (and the equivalent orthogonal vector). Below \Tone, the weak orthorhombicity along with the relatively large electron-phonon coupling \cite{Yao:2006} then tips the balance, and \vQ$_1$ aligns along $c^\ast$ rather than $a^\ast$. Finally, upon further lowering the temperature below $T_{\rm CDW2}$ $\vQ_2$ aligns along $a^\ast$ since the Fermi surface along the $c^\ast$ direction is already fully gapped by the first transition \cite{Ru:2008a,Moore:2010}.

Therefore, we identify two cooperating effects determining the overall selection of the ordering vector: (i) The system gains energy by gapping the band degeneracy points on the Fermi surface where the Raman selection rules indicate substantial fluctuations  with fourfold symmetry above \Tone. (ii) Since the electron-phonon coupling vertex is proportional to the Raman vertex for stress phonons \cite{Shastry:1990,Miller:1967}, both quantities are enhanced near band degeneracies. For small hybridization and an electron-phonon coupling strength of $\lambda>0.5$ the focussing effect may even be the most relevant contribution to the phonon renormalization, and thus the CDW formation, while it is only a correction for larger $V_{xz}$. Hence, while electron-phonon coupling is known to be important in CDW systems \cite{Mcmillan:1977,Varma:1983,Yao:2006}, we identify on a microscopic basis the focussing effect to be a more generic paradigm for multiband materials. As a future outlook, it seems particularly interesting to explore this novel scenario in the proximity of superconductivity, eventually competing or coexisting with CDW order. In fact, superconductivity at approximately $2\,\kelvin$ appears in some of the rare earth tri-tellurides if the CDW is suppressed by applied pressure \cite{Hamlin:2009,Maple:2012}. Moreover, it would be intriguing to address in a wider context the effects of band hybridization in materials such as the iron-based superconductors \cite{Mazin:2010a}, in which density-wave order and superconductivity interplay on a microscopic scale.

\begin{acknowledgments}
We benefited from discussions with B. Moritz, R.\,G.~Moore, and B. Muschler and thank T. B\"ohm for assistance. R.H. thanks the Stanford Institute for Materials and Energy Sciences, where part of the paper was completed, for its hospitality. A.F.K. and T.P.D. thank the Walther Meissner Institut for its hospitality. Financial support from Deutsche Forschungsgemeinschaft Grant HA 2071/5-1 and the Collaborative Research Center TRR~80 is gratefully acknowledged. L.D. acknowledges support by the Swiss National Foundation for the Scientific Research within the pool, "Materials with Novel Electronic Properties" of the National Centres of Competence in Research network. E.A.N., A.F.K., T.P.D., J.-H.C., J.G.A., and I.R.F. acknowledge support from US Department of Energy, Basic Energy Sciences, Materials Sciences and Engineering Division Contract
No. DE-AC02-76SF00515.
\end{acknowledgments}


\begin{appendix}

\section{Samples and experimental technique}
\label{appendixA}

Well characterized single crystals of ErTe$_3$ were grown by slow cooling of a binary melt as described elsewhere \cite{Ru:2006,Ru:2008a}. ErTe$_3$ is a particularly well ordered system. In the $a$-$c$ plane, the resistivity varies only slowly above \Tone~ and is very small in the $T=0$ limit \cite{Ru:2008a}. The crystals were cleaved before being mounted into the cryostat.

The imaginary part $R\chi^{\prime\prime}_{i,s}(\omega,T)$ of the Raman response is measured for various polarization combinations of incoming and scattered photons (${\bf e}_{i}$ and ${\bf e}_{s}$) referred to as $aa$, $cc$, and $ac$ using Porto notation. Symbolic representations by two arrows in the Te plane are shown along with the spectra. Usually, more than one symmetry component is projected out at a given polarization ${\bf e}_{i}, {\bf e}_{s}$. The pure symmetries correspond to specific eigenvectors in the case of phonons and to separate regions in the Brillouin zone for electron-hole excitations \cite{Devereaux:2007}. In this publication we show predominantly spectra with $aa$ and $cc$ polarizations which comprise $A_{1g}$ and $B_{1g}$ symmetry components in a tetragonal lattice and the $A_{g}$ symmetry on an orthorhombic lattice. In either case, the response has C$_2$ symmetry.

For the experiments we made use of a solid state laser emitting at 532.3\,nm (KLASTECH SCHERZO\,300) for excitation. The absorbed laser power ranged from 1 to 2\,mW to keep the local heating below 5 K in the $50\times100\,\mu$m$^2$  sized focus. The spectra were measured with a resolution of 2.5 \wn ~at low energy and 20 \wn ~at high energy. The Raman response $R\chi^{\prime\prime}$ is then obtained by dividing the measured spectra by the thermal Bose factor $\{1+n(\omega,T)\}=[1-e^{-\hbar\omega/k_BT}]^{-1}$. $R$ is a constant which absorbs experimental factors and takes care of the units.

\section{Theory}
\label{appendixB}

As noted in the main text, the anisotropic electron-phonon coupling vertex must be included in the electronic susceptibility.  For the case of stress phonons, the electron-phonon coupling and Raman vertices are related via $g_{\bf k}=g\gamma_{\bf k}$ \cite{Shastry:1990,Keck:1976}, where $g$ sets the strength of the overall electron-phonon interaction.  We therefore define the projected electronic susceptibility as
\begin{equation}
  \chi_P({\bf q},\Omega)=2 \sum_{{\bf k}} \gamma_{\bf k}^+ \gamma_{\bf k}^- \frac{f(\epsilon_{{\bf k}+{\bf q}/2}^+)-f(\epsilon_{{\bf k}-{\bf q}/2}^-)}{\Omega + i\delta+\epsilon_{{\bf k}+{\bf q}/2}^+-\epsilon_{{\bf k}-{\bf q}/2}^-}
  \label{eq:chi_P}
\end{equation}
where $\epsilon_{\bf k}^\pm$ are the two bands and $\gamma_{\bf k}^\pm=\gamma_{aa}^\pm+\gamma_{cc}^\pm$ are the fully symmetric effective mass vertices derived in the SI. %
We consider only interband contributions which are generally accentuated by nesting as shown in Fig.\;\ref{fig:structure}\emph{C}. Here we wish to explore how nesting and anisotropic electron-phonon coupling conspire to ultimately select the experimentally observed ordering vector $\vQ_1$.
$\chi_P({\bf q},\Omega)$ leads to a significant phonon-softening at the wavevector $\vQ_1$ connecting the corners of the Fermi surface, where the band degeneracy is lifted. The effect is very sensitive to the hybridization parameter $V_{xz}$. Results for a set of hybridization parameters are shown in Fig.\;\ref{fig:phonon}.


\section{Supporting information}
\label{appendixC}

\subsection{Raman scattering and band structure}

For the case of non-resonant light scattering, the Raman cross section in the limit of small momentum transfer is given in terms of correlation functions of an effective charge density \cite{Devereaux:2007},
\begin{equation}
  \tilde{\rho}=\sum_{\nu,\mathbf{k},\sigma}\gamma_{\nu}(\mathbf{k},\omega_{i},\omega_{s}) c_{\nu,\mathbf{k},\sigma}^{\dagger}c_{\nu,\mathbf{k},\sigma},
  \label{eq:rho}
\end{equation}
where $c_{\nu,\mathbf{k},\sigma}^{\dagger}$ ($c_{\nu,\mathbf{k},\sigma}$) creates (removes) an electron with momentum $\mathbf{k}$ and spin $\sigma$ in (from) band $\nu$. In principle, the scattering amplitude $\gamma_\nu$  depends on both the incoming (scattered)  light polarizations ${\mathbf{e}^{i(s)}}$ and frequencies  $\omega_{i(s)}$. However, in the limit where $\omega_{i(s)}$ are much smaller than any relevant interband transition frequency, the scattering amplitude simplifies to the well known tensor of the inverse effective mass,
\begin{equation}
  \gamma_{\nu}(\mathbf{k},\omega_{i,s}\rightarrow0)=\frac{1}{\hbar^2}\sum_{\alpha,\beta
  =x,y,z}e_{\alpha}^{i}\frac{\partial^{2}\epsilon^{(\nu)}(\mathbf{k})}{\partial
  k_{\alpha}\partial k_{\beta}}e_{\beta}^{s}.
  \label{eq:gamma}
\end{equation}
Thus the curvature of the bands and the light polarization orientations determine which carriers are involved in light scattering in different bands and regions of the Brillouin zone.

The tri-telluride band structure for the two bands cutting the Fermi level is well described by a tight binding model that only includes the Te $p_x$ and $p_z$ orbitals \cite{Brouet:2008}. All expressions in this paper are given for coordinates in the unit cell defined by the square lattice of Te atoms (Fig.\;1\emph{B}).

The real space Hamiltonian  can be expressed as
\begin{eqnarray}
  H\! &=&\!-V_{p\sigma} \sum_{m,n}\left[c_{m,n+1(px)}^\dagger c_{m,n(px)}\!+\!c_{m+1,n(pz)}^\dagger c_{m,n(pz)}\right]\nonumber \\
  &-&\!\!\!V_{p\pi}\sum_{m,n}\left[c_{m+1,n(px)}^\dagger c_{m,n(px)}\!+\!c_{m,n+1(pz)}^\dagger c_{m,n(pz)}\right]\nonumber\\
  &-&\!\!\!V_{xx}\!\sum_{m,n,\alpha}\!\left[ c_{m+1,n+1(\alpha)}^\dagger c_{m,n(\alpha)}\!+\!c_{m-1,n+1 (\alpha)}^\dagger c_{m,n(\alpha)}\right]\nonumber\\
  &-&\!\!\!V_{xz}\!\sum_{m,n}\!\left[c_{m+1,n+1(px)}^\dagger c_{m,n(pz)}\!+\!c_{m+1,n+1(pz)}^\dagger c_{m,n(px)}\right]\nonumber\\
  &+&\!\!\!V_{xz}\!\sum_{m,n} \!\left[c_{m-1,n+1(px)}^\dagger c_{m,n(pz)}\!+\!c_{m-1,n+1(px)}^\dagger c_{m,n(pz)}\right] \nonumber\\
  &-&\!\!\!\mu\sum_{m,n,\alpha}c_{m,n,\alpha}^\dagger c_{m,n,\alpha} \!+\! h.c.
   \label{eq:Hamiltonian_tight}
\end{eqnarray}
where $c^\dagger_{m,n(\alpha)}$ creates an electron at site $(m, n)$ in orbital $\alpha =p_x, p_z$. It includes both nearest and next-nearest neighbor hopping matrix elements of one tellurium plane, as denoted in Fig.\;1\emph{B}. Fourier transforming Eq.~(\ref{eq:Hamiltonian_tight}) leads to the two bands $\epsilon_{\bf k}^{(\nu=\pm)}$, where $\nu=+$ and $\nu=-$ label the electron and the hole band, respectively,
\begin{equation}
  \epsilon_{\bf k}^{\pm} = \frac{1}{2} \left[h^{11}_{\vk}+h^{22}_{\vk} \pm \sqrt{(h^{11}_{\vk}-h^{22}_{\vk} )^2+4(h^{12}_{\vk})^2} \right].
\label{eq:bands}
\end{equation}
With the Te-Te distance set to unity, the energies $h^{i,j}$ read
\begin{eqnarray}
  h_{\vk}^{11} \! &=& \! -2V_{p\pi}\cos k_z \! - \! 2V_{p\sigma}\cos k_x \! - \! 4V_{xx}\cos k_z\cos k_x \! - \! \mu, \nonumber\\
  h_{\vk}^{22} \! &=& \! -2V_{p\pi}\cos k_x \! - \! 2V_{p\sigma}\cos k_z \! - \! 4V_{xx}\cos k_z\cos k_x \! - \! \mu, \nonumber\\
  h_{\vk}^{12} \! &=& \! -4V_{xz} \sin k_z \sin k_x.
\end{eqnarray}

\begin{figure*}
  \centering
  \includegraphics[width=0.5\textwidth]{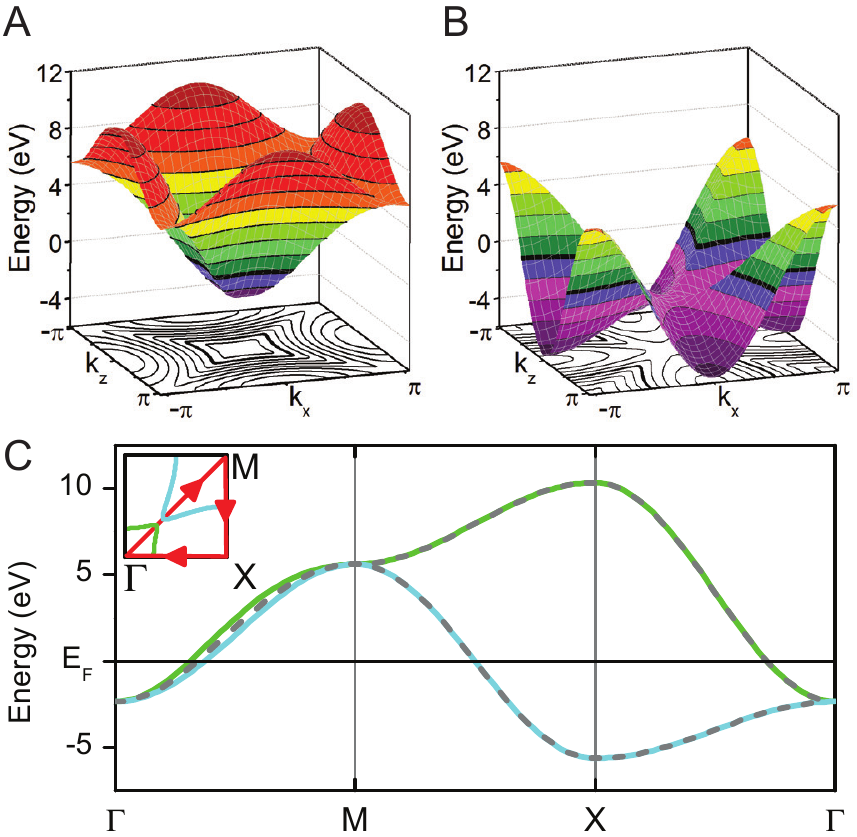}
  \caption{Band structure of ErTe$_3$ including diagonal hopping. Panels (\emph{A}) and (\emph{B}) show the electron and hole band, respectively, using $V_{xx}=0.09$\,eV and a hybridization of $V_{xz}=0.12$\,eV. Both the 2D band structure and the constant energy contours are plotted. The sharp corners result from the hybridization $V_{xz}$ and lead to the huge enhancement of the Raman vertices shown in Fig.\;5\emph{D}-\emph{I}. Panel (\emph{C}) shows the dispersion along the line $\Gamma - M - X - \Gamma$ as indicated in the inset before (dashed lines) and after (full lines) the hybridization $V_{xz}$ has been turned on.
  }
  \label{fig:band}
\end{figure*}

Using the experimental distances between the rare-earth and the Te ions  of 4.34 and $4.34/\sqrt{2}$\,\AA, respectively, the hopping matrix elements are $V_{p\sigma} = 2.99, V_{p\pi} =-1$\,eV. The diagonal hopping integrals are chosen to be $V_{xx}=0.09, V_{xz}=0.12$\,eV. This parameter choice is motivated in order to most closely match the Fermi surfaces derived from angle-resolved photoemission spectroscopy (ARPES)~\cite{Moore:2010}. The chemical potential is chosen to yield a filling of 1.6 electrons/band. Any small interaction (here we  focus on inter-orbital hybridization $V_{xz}$) leads to sharp corners in the electron and hole band, displayed separately in Fig.\;\ref{fig:band}\emph{A} and \emph{B}, respectively. This hybridization thus lifts the band degeneracy along the $\Gamma-M$ Brillouin zone cut with respect to the $V_{xz}=0$ case, as shown in Fig.\;\ref{fig:band}\emph{C}.

Finally, we shall note that a more accurately derived tight-binding representation would require the inclusion of further hopping matrix elements that could affect the overall curvature of the bands. Furthermore, another larger unit cell taking into account the Te double layers is usually considered for addressing the ARPES results \cite{Brouet:2008}. While these latter issues would allow a fine tuning of the calculations, we trust that this would not alter the main conclusion of our work, effectively based on a simplified tight-binding approach for a small unit cell.

\subsection{Weak-coupling Raman response}

The Hamiltonian for a system with two CDWs can be expressed in Nambu notation as $H_{\textrm{CDW}}=H({\bf Q}_1,\Delta_1)+H({\bf Q}_2,\Delta_2)$, where
\begin{equation}
  H({\bf Q},\Delta)=\sum_{{\bf k}\sigma} \psi_{{\bf k}\sigma}^\dagger \hat{\Lambda}_{\bf k}(\Delta) \psi_{{\bf k}\sigma}.
  \label{eq:Hamiltonian}
\end{equation}
This Hamiltonian is defined in terms of the spinor $\psi_{{\bf k}\sigma}=(c_{+,{\bf k}-{\bf Q},\sigma},c_{+,{\bf k},\sigma}, c_{+,{\bf k}+{\bf Q},\sigma},c_{-,{\bf k}-{\bf Q},\sigma},c_{-,{\bf k},\sigma}, c_{-,{\bf k}+{\bf Q},\sigma})$ where $c_{\pm,{\bf k},\sigma}$ destroys an electron in band $\epsilon_{\bf k}^\pm$ with momentum \textbf{k} and spin $\sigma$. The matrix $\hat{\Lambda}_{\bf k}$ is  given by
\begin{eqnarray}
  \hat{\Lambda}_{\bf k} (\Delta)=
  \left (
  \begin{array}{cccccc}
  \epsilon_{{\bf k}-{\bf Q}}^{+} & \Delta & 0 & 0 & \Delta & 0 \\
  \Delta & \epsilon_{{\bf k}}^{+} & \Delta & \Delta & 0 & \Delta  \\
  0 & \Delta & \epsilon_{{\bf k}+{\bf Q}}^{+} &0 & \Delta & 0 \\
  0 & \Delta & 0 & \epsilon_{{\bf k}-{\bf Q}}^{-} & \Delta & 0 \\
  \Delta & 0 &\Delta & \Delta & \epsilon_{{\bf k}}^{-} & \Delta \\
  0 & \Delta & 0 & 0 & \Delta & \epsilon_{{\bf k}+{\bf Q}}^{-}
  \end{array}
  \right )
  \label{eq:energy}
\end{eqnarray}
where the diagonal elements correspond to the quasiparticle energies at ${\bf k}-{\bf Q}$, ${\bf k}$, and ${\bf k}+{\bf Q}$, and $\Delta$ is the CDW gap. For \vQ\, and $\Delta$ we use ${\bf Q}_1 = (2.16, 0, 2.16)/a$,  ${\bf Q}_2 = (2.16, 0, -2.16)/a$, and $\Delta_1 \!=\! 1400 $\,cm$^{-1}$ and $\Delta_2 \!=\! 400 $\,cm$^{-1}$, respectively. The expansion is truncated after the first harmonic although higher order terms $n{\bf Q}$ should be included for the incommensurability of ${\bf Q}_{1,2}$ \cite{Brouet:2008}.

Computing the matrix Green's function $\hat{G}({\bf k},i\omega_n)=(i\omega_n-\hat{\Lambda}_{\bf k})^{-1}$ and taking into account the Raman vertices leads to the Raman response \cite{Vanyolos:2005,Devereaux:2007}
\begin{eqnarray}
  \chi_{\Gamma,\Gamma}({\bf q}\!\!&=&\!\!0,i\Omega_n)= \label{eq:response} \\
  &-&\!\!\frac{2}{V\beta} \sum_{{\bf k},i\omega_n} \textrm{Tr}[\hat{\Gamma}_{\bf k}\hat{G}({\bf k},i\omega_n)\hat{\Gamma}_{\bf k}\hat{G}({\bf k},i\omega_n+i\Omega_n)]. \nonumber
\end{eqnarray}
Here, $i\Omega_n$ and $i\omega_n$ are the external and internal Matsubara frequencies, respectively, $\beta=(k_BT)^{-1}$, and Tr denotes the trace.

The Raman vertex $\hat{\Gamma}_{\bf k}$ is
\begin{eqnarray}
  \hat{\Gamma}_{\bf k} =
  \left(
  \begin{array}{cccccc}
  \gamma_{{\bf k}-{\bf Q}}^{+} & 0 & 0 & 0 & 0 & 0 \\
  0 & \gamma_{\bf k}^{+} & 0 & 0 & 0 & 0 \\
  0 & 0 & \gamma_{{\bf k}+{\bf Q}}^{+} & 0 & 0 & 0 \\
  0 & 0 & 0 & \gamma_{{\bf k}-{\bf Q}}^{-} & 0 & 0 \\
  0 & 0 & 0 & 0 & \gamma_{\bf k}^{-} & 0 \\
  0 & 0 & 0 & 0 & 0 & \gamma_{{\bf k}+{\bf Q}}^{-}
  \end{array}
  \right)
  \label{eq:Gamma}
\end{eqnarray}
with $\gamma^{\pm}$ labeling the effective mass vertices [Eq.~(\ref{eq:gamma})] derived from $\epsilon_{\vk}^{\pm}$ [Eq.~(\ref{eq:bands})]. Eq.~(\ref{eq:response}) can be evaluated by diagonalizing the Green's functions before calculating the trace, which simplifies the analytic structure of the expression and allows us to express the response as a sum of Lindhard-like terms.


\begin{figure*}
  \centering
  \includegraphics[width=0.45\textwidth]{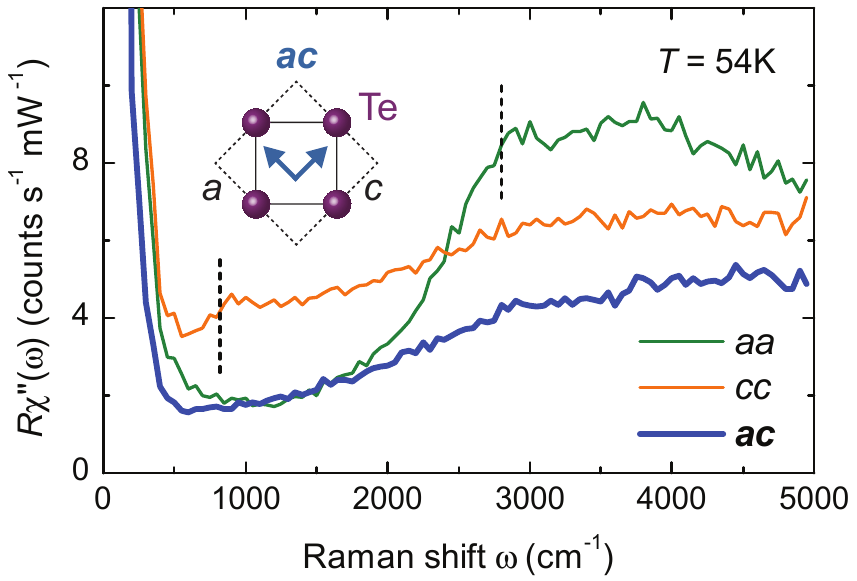}
  \caption{Electronic Raman response of \ErTe\,for parallel and crossed polarizations. The inset sketches the incoming and outgoing photon polarizations for the $ac$ configuration (defined in the methods section). At $T=54\,\kelvin$ both CDWs are present in the sample but are clearly detected only in the $aa$ and $cc$ polarization.
  }
  \label{fig:nogap}
\end{figure*}


The Raman response calculated with Eq.~(\ref{eq:response}) is shown in Fig.\;5. In Fig.\;\ref{fig:nogap} we show the experimental response for parallel polarizations $aa$ and $cc$ in comparison to that for crossed polarizations $ac$. In accordance with the theoretical prediction the $ac$ response shows very little or no signatures of the gaps.

\subsection{Effect of near band degeneracies on the phonon spectrum}

We recall  that in a conventional electron-phonon coupling scenario, a divergence in the electronic susceptibility is translated into the phonon dispersion via the electron-phonon coupling \cite{Kohn:1959}. With increasing interaction $\lambda$, the phonon frequency $\omega_{\bf q}$ of a branch is first renormalized in a wide range of momenta, then softens and finally exhibits a discontinuous derivative at wave vectors \vQ~ that nest the Fermi surface. A phonon that softens to zero frequency at a certain wave vector thus signifies the formation of a static lattice distortion, or CDW with that ordering vector.

The renormalized phonon frequency $\tilde{\omega}_{\bf q}$  is given by~\cite{Gruner:1994}
\begin{equation}
  \tilde{\omega}_{\bf q}^2=\omega_{\bf q}^2[1-\lambda({\bf q},\omega_{\bf q})],
  \label{eq:w-R}
\end{equation}
where
\begin{equation}
  \lambda({\bf q},\omega_{\bf q})=\frac{2g^2 \chi'({\bf q},\omega_{\bf q})}{\omega_{\bf q}}.
  \label{eq:lambda-q}
\end{equation}
$\chi'({\bf q},\omega_{\bf q})$ is the electronic susceptibility. $g$ in Eq.~(\ref{eq:lambda-q}) sets the overall strength of the electron-phonon interaction. The phonon renormalization is expected to be quite relevant in the tri-tellurides, since Raman experiments have established a strong connection between the CDW gap and lattice distortions \cite{Lavagnini:2008}.


\begin{figure*}
  \centering
  \includegraphics[width=0.45\textwidth]{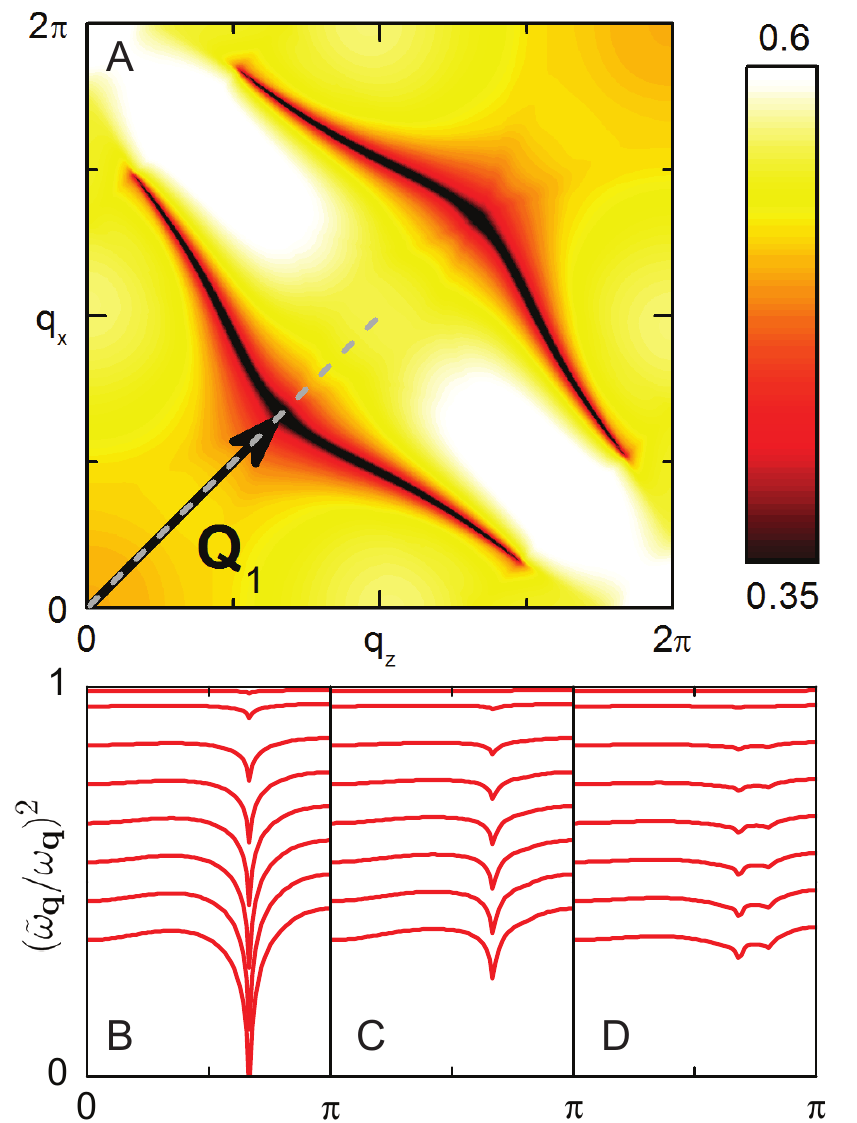}
  \caption{Renormalized phonon dispersion relation. (\emph{A}) Renormalized phonon frequency $(\tilde{\omega}_{\bf q}/\omega_{\bf q})^2$ with electron-phonon coupling $\lambda_{{\bf q}=0}=0.5$ and diagonal hopping terms set for clarity to 10 times smaller values than those used in Figs.\,1\emph{D} and 5. The arrow represents the ordering wavevector $\vQ_1$ as determined  experimentally. (\emph{B}-\emph{D}), Cuts along the line $q_x=q_z$ (dashed line in panel \emph{A}) for various values of the electron-phonon coupling starting at $\lambda = 0.01$ (top curve) and then ranging between $\lambda=0.05$ and $\lambda=0.65$ (bottom curve) with incremental steps of 0.1. Each panel is for a different value of the hybridization parameters: (\emph{B}) $V_{xx}/10$, $V_{xz}/10$;  (\emph{C}) $V_{xx}/4$, $V_{xz}/4$; and (\emph{D}) $V_{xx}$, $V_{xz}$.
  }
  \label{fig:phonon}
\end{figure*}


We now examine the effect of charge fluctuations on the phonon renormalization. We start with the original proposal by Kohn, \cite{Kohn:1959} where the Lindhard susceptibility $\chi_L$ directly couples to the phonon propagator with a bare charge vertex,
\begin{equation}
  \chi_L({\bf q},\Omega) = 2 \sum_{\bf k\alpha\beta}
  \frac{ f(\epsilon^\alpha_{\bf k+q}) - f(\epsilon^\beta_{\bf k})}
  {\Omega+i\delta + \epsilon^\alpha_{\bf k+q} - \epsilon^\beta_{\bf k}}.
  \label{eq:chi_L}
\end{equation}
Here, $\alpha$ and $\beta$ label the bands. The effect of the normal charge susceptibility $\chi_L$ has been studied in some detail by several authors \cite{Yao:2006,Johannes:2008}. However, within this picture, the selection of the ordering vector is not pronounced; the maxima in the susceptibility are almost equally high for a large manifold of ordering vectors (Fig.\;6\emph{A}), supporting the conclusion that another mechanism is responsible for finally selecting the experimentally observed ordering vector $\vQ_1$.

We thus consider a generic coupling to stress phonons where the electron-phonon coupling vertex is determined by the inverse effective mass tensor, which is related to the Raman vertex via $g_{\bf k}=g\gamma_{\bf k}$ \cite{Shastry:1990,Keck:1976}. As a result, in regions of strong band curvature the electron-phonon coupling is  large. We consider only interband contributions which are generally accentuated by nesting as shown in Fig.\;1\emph{C}. Using the mass density operator [Eq.~(\ref{eq:rho})] we define the projected susceptibility $\chi_P$ for inter-band scattering transitions as
\begin{equation}
  \chi_P({\bf q},\Omega)=2 \sum_{{\bf k}} \gamma_{\bf k}^+ \gamma_{\bf k}^- \frac{f(\epsilon_{{\bf k}+{\bf q}/2}^+)-f(\epsilon_{{\bf k}-{\bf q}/2}^-)}{\Omega+i\delta+\epsilon_{{\bf k}+{\bf q}/2}^+-\epsilon_{{\bf k}-{\bf q}/2}^-}
  \label{eq:chi_P}
\end{equation}
where $\epsilon_{\bf k}^\pm$ are the two bands defined in Eq.~(\ref{eq:bands}) and $\gamma_{\bf k}^\pm=\gamma_{aa}^\pm+\gamma_{cc}^\pm$ are the fully symmetric effective mass vertices derived from these bands [Eq.~(\ref{eq:gamma})]. Note that this quantity is only $C_4$ symmetric in the larger momentum range ${\bf q}_{x(z)}=(0, 4\pi)$ and vanishes for $V_{xz}=0$. We use this susceptibility in our calculation  of the renormalized phonon frequency [Eqs.~(\ref{eq:w-R}) and (\ref{eq:lambda-q})]. Here we have taken the bare phonon frequency $\omega_{\bf q}$ to be independent of momentum, hence $\omega_{\bf q} \equiv \omega_{2{\bf q}_F}$, as any mild momentum dependence will not affect the general behavior near ${\bf Q}_1$.

The phonon frequency renormalized by $\chi_P({\bf q},\Omega)$ is shown in Fig.\;\ref{fig:phonon}\emph{A} for small hybridization (i.e., with values of $V_{xx}$ and $V_{xz}$ 10 times smaller than those employed in Figs.\,1\emph{D} and 5) in order to emphasize the effect. Panels \emph{B}-\emph{D} further display the phonon softening along the $(0,0)-(\pi,\pi)$ Brillouin zone cut (dashed line in panel \emph{A}) as a function of $V_{xz}$ and $\lambda$. It is evident that the phonon frequency exhibits a significant softening at the wavevector $\vQ_1$ connecting the corners of the Fermi surface, where the band degeneracy is lifted by hybridization.
This predicted phonon renormalization remains to be experimentally confirmed by neutron or inelastic x-ray scattering (RIXS), since Raman scattering is limited to small momentum transfer $\vq \ll \vQ_1$. The focusing effect of the large curvature of the energy bands given by the mass tensor is very sensitive to the hybridization parameter $V_{xz}$ (Fig.\;\ref{fig:phonon}\emph{B}-\emph{D}). Nonetheless, our study has a model character pointing out the relevance of the projected susceptibility for interband scattering $\chi_P$. The exact strength of the enhancement due to the $\chi_P$ at the experimentally observed ordering vector $\vQ_1$ depends critically on the dispersion over the whole Brillouin zone \cite{Johannes:2008} which is left to a future study.

\end{appendix}

\end{document}